# Spectral singularities and threshold gain of a slab laser under illumination of a focused Gaussian beam

**Mehrdad Bavaghar, Rasoul Aalipour*, and Kazem Jamshidi-Ghaleh**

*Department of Physics, Azarbaijan Shahid Madani University, Tabriz 53714-161, Iran*
*aalipour@azaruniv.ac.ir

**Abstract:** We study spectral singularities of an infinite planar slab of homogenous optically active material in focus of a thin lens under illumination of a Gaussian beam. We describe the field distribution of the Gaussian beam under this configuration as a plane wave propagated near the optical axis which it's phase and amplitude vary with distance from center of the beam. Based on this approximation, we carry out the transfer matrix for the slab. We explore the consequences of this configuration on determining the threshold gain of the active medium and tuning the resonance frequencies related to spectral singularities. We show that the spectral singularities and the threshold gain besides that vary with distance from center of the Gaussian beam, also they change with relative aperture of the focusing lens. As a result, using a thin lens with higher relative aperture, the spectral singularities corresponding resonances shift to the higher frequencies (lower wavelengths). Numerical results confirm the theoretical findings.

## 1. Introduction

Spectral singularities are certain points of the continuous spectrum of Schrodinger operator for a complex potential with real scattering energies at which the transmission and reflection amplitudes of the potential diverge. Physically they correspond to scattering states that behave like zero-width resonances [1]. They draw the attention of mathematicians because they are responsible for a number of mathematical peculiarities that can never arise for Hermitian operators and fascinated physicists because of their optical applications. Spectral singularity was discovered by Naimark and then studied by some mathematicians and physicists [2–5]. Applications of the spectral singularities in optical systems were introduced by Mostafazadeh [6]. He show that a slab involving optical gain material begins emitting purely outgoing coherent waves at resonance frequencies related to the spectral singularities, i.e., it acts as a slab laser. This observation has provided sufficient motivation for the continuous study of spectral singularities and their applications [7–16, 18, 19, 21–25].
Among the notable applications of this approach, those focusing on the structure of incident light and variety of the optical setups, are of interest to us [17, 20, 26, 27]. In recent relevant studies, the electromagnetic wave incident on the optically active medium is a plane wave, for which the transfer matrix for getting the transmission and reflection spectra of the medium is easily calculated. On one hand, the field distribution of most lasers is Gaussian, therefore, it is worthwhile to investigate the spectral singularities of the active medium under illumination of a Gaussian beam. Also, we know that the amplitude and phase of a Gaussian beam is a quadratic function of the radius of beam, so makes it difficult to use the transfer matrix method. However, by using proper optical setups, one can use the transfer matrix method for the Gaussian beam. In this paper, we consider the spectral singularities of a slab laser in focus of a thin lens under illumination of a Gaussian beam, for which the transfer matrix method is applicable. The paper is organized as follows: In Sec. 2, we describe the field distribution of a Gaussian beam in focus of a thin lens. In Sec. 3, the transmission and reflection coefficients of a Gaussian beam from the laser slab are calculated using the transfer matrix method and the associated spectral singularities are obtained. We express the equation related to threshold gain of the slab laser under illumination of a focused Gaussian beam in Sec. 4. In Sec. 5, we account the dispersion relations for the equations governing the spectral singularities and the threshold gain. Finally, we report the numerical results in Sec. 6.

## 2. Illumination an infinite planner slab by a focused Gaussian beam

In Fig. 1, an infinite planar slab of thickness $L$ is installed in back focal plane $(x, y)$ of a convergent thin lens. Suppose that the slab contains a homogeneous material with the complex refractive index, $n$. A Gaussian light beam propagating along the $z$-axis is focused on entrance face of the slab through the lens. The transverse component of the electric field describing such an electromagnetic wave can be written as follows [28]

$$\overrightarrow{\mathrm{E}}(\mathrm{r}, t) = \psi(\mathrm{r}) e^{-i\omega t} \hat{e}_x, \tag{1}$$

where $\mathrm{r} := (x, y, z)$, $\hat{e}_x$ is the unit vector along the positive $x$-axis, $\omega$ is the frequency of the light, and $\psi(\mathrm{r})$ is a continuously differentiable scalar wave function. Inserting this component of the field in the familiar Helmholtz equation leads to the following time-independent Schrodinger equation:

$$-\nabla^2\psi(\mathrm{r}) + v(z)\psi(\mathrm{r}) = k^2\psi(\mathrm{r}), \tag{2}$$

with the complex potential $v(z) = k^2(1 - \mathfrak{z}^2)$, where $k := \frac{\omega}{c}$ stands for the wave number, $c$ is the speed of light in vacuum, and

$$\mathfrak{z} := \begin{cases} n & |z| \le \frac{L}{2} \\ 1 & |z| > \frac{L}{2} \end{cases}. \tag{3}$$

The main characteristic of a Gaussian beam is that it concentrates near the propagation direction ($z$-axis), allowing it to be described by a scalar wave of the form [29]

$$\psi(\mathrm{r}) = \acute{\psi}(\mathrm{r}) e^{i\mathfrak{z}kz} \tag{4}$$

i.e., the wave does not propagate in the $x$ or $y$ direction and $\acute{\psi}(\mathrm{r})$ is slowly varying with $z$. Substituting Eq.(4) into Eq. (2) and ignoring $\frac{\partial^2\acute{\psi}}{\partial z^2}$ leads to the following paraxial wave equation:

$$\frac{\partial^2\acute{\psi}}{\partial x^2} + \frac{\partial^2\acute{\psi}}{\partial y^2} + 2i\mathfrak{z}k\frac{\partial\acute{\psi}}{\partial z} = 0. \tag{5}$$

It's solution gives the field distribution of the Gaussian beam as follows [29]:

$$\psi(\mathrm{r}) = E_0 \frac{w_0}{w(z)} \exp[-\frac{r^2}{w^2(z)}] e^{i\mathfrak{z}kz} \exp[i\phi(z)] \exp[-\frac{i\mathfrak{z}kr^2}{2R(z)}], \tag{6}$$

where $r^2 = x^2 + y^2$, $R(z) = z[1 + (z_0/z)^2]$, $w(z) = w_0\sqrt{1 + (z/z_0)^2}$, and $\phi(z) = tan^{-1}(z/z_0)$ in which $w_0$ and $z_0$ are the minimum beam waist and Rayleigh range, respectively. The Rayleigh range can be written in terms of the minimum beam waist as $z_0 = \frac{\pi w_0^2}{\lambda}$ [29].
When the Gaussian beam is followed through a lens of diameter $D$ and focal length $f$, the minimum beam waist lies to the right of the lens, a distance $f$ from the lens. The minimum beam waist and the Rayleigh range are given by [29]

$$w_0 \approx 2\lambda(f/D) \tag{7}$$

and

$$z_0 \approx 4\pi\lambda(f/D)^2, \tag{8}$$

where the ratio $f/D$ is defined as the relative aperture of lens [30].

According to Fig. 1, we choose the origin of the coordinates system at back focal plane of the lens, where installed the slab. Suppose that the thickness of the slab is very small in respect to the Rayleigh range, $L \ll z_0$. Therefore, we will be allowed to use the approximation $z \ll z_0$ for the Gaussian beam parameters as follows:

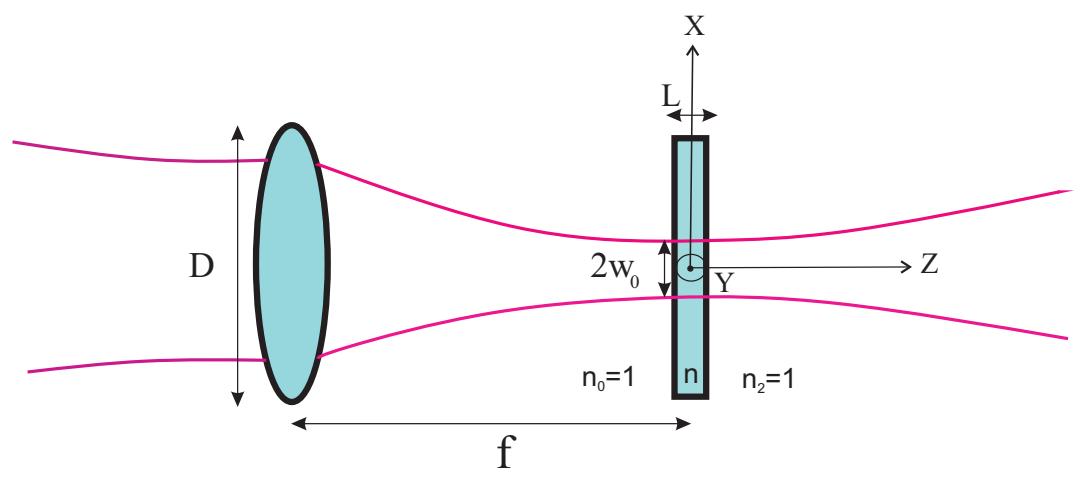

Fig. 1. Sketch of an infinite planar slab of thickness $L$ and complex refractive index $n$ in focus of a convergent thin lens of diameter $D$ and focal length $f$. The slab is illuminated by a Gaussian beam through the lens.

$$R(z) \approx \frac{z_0^2}{z} \qquad w(z) \approx w_0 \qquad \phi(z) \approx 0. \tag{9}$$

Substituting these relations in Eq. (6) leads to the following equation for the field distribution of the Gaussian beam in $z \ll z_0$ approximation:

$$\psi(\mathrm{r}) = A_i(r)e^{i\mathfrak{z}k\xi(r)z}, \tag{10}$$

where

$$A_i(r) = E_i e^{-\frac{r^2}{w_0^2}}, \tag{11}$$

and

$$\xi(r) = [1 + \frac{r^2}{2z_0^2}]. \tag{12}$$

The scalar wave function in Eq. (10) describes a plane wave which it's phase and amplitude are $r$-dependent [31]. Using these results, the general solution of the Schrodinger equation, Eq. (2), can be expressed as follows:

$$\psi(r,z) = \begin{cases} A_0(r)e^{ik\xi(r)z} + B_0(r)e^{-ik\xi(r)z} & \text{for} \quad z < -\frac{L}{2} \\ A_1(r)e^{ink\xi(r)z} + B_1(r)e^{-ink\xi(r)z} & \text{for} \quad |z| \leq \frac{L}{2} \\ A_2(r)e^{ik\xi(r)z} + B_2(r)e^{-ik\xi(r)z} & \text{for} \quad z > \frac{L}{2}, \end{cases} \tag{13}$$

where $A_i$ and $B_i$, with $i = 0, 1, 2$ are $r$-dependent forward and backward wave amplitudes according to the Eq. (11).

## 3. Transfer matrix and spectral singularities

The transfer matrix for the system we consider is the $2 \times 2$ matrix $\mathbf{M}$ satisfying

$$\begin{pmatrix} A_2 \\ B_2 \end{pmatrix} = \mathbf{M} \begin{pmatrix} A_0 \\ B_0 \end{pmatrix}. \tag{14}$$

We explore the transfer matrix by using the boundary conditions at $z = -\frac{L}{2}$ and $z = \frac{L}{2}$ based on continuity of the tangential components of electric and magnetic fields. The transfer matrix after some straightforward calculations is obtained as follows:

$$\mathbf{M} = \frac{1}{4n}\begin{pmatrix} e^{-ikn\xi(r)L}F(n, -\frac{kL}{2}) & 2i(n^2-1)\sin[kn\xi(r)L] \\ -2i(n^2-1)\sin[kn\xi(r)L] & e^{ikn\xi(r)L}F(n, \frac{kL}{2}) \end{pmatrix}, \tag{15}$$

where for all $p, q \in C$

$$F(p,q) = e^{-2ipq}(1+p)^2 - e^{2ipq}(1-p)^2. \tag{16}$$

From Eq. (15) it is clear that the transfer matrix elements depend on the transverse coordinate $r$. For a case that we consider the center of beam, i.e., $r = 0$, Eq. (15) turns to the common transfer matrix for the slab laser under illumination of a plane wave.
The left and right reflection and transmission amplitudes are given by [1]

$$R_l = -\frac{M_{21}}{M_{22}} \quad R_r = \frac{M_{12}}{M_{22}} \quad T_l = \frac{\det M}{M_{22}} \quad T_r = \frac{1}{M_{22}}, \tag{17}$$

where $M_{ij}$ $i, j = 1, 2$ are the elements of transfer matrix $M$ and det stands for determinant of the matrix which one can show that $\det M = 1$. Inserting transfer matrix elements from Eq. (15) in Eq. (17) yields the following relations:

$$R_l = R_r = e^{-ik\xi(r)L}\frac{\sqrt{\mathrm{R}}(1 - e^{2ik\xi(r)nL})}{1 - \mathrm{R}e^{2ik\xi(r)nL}} \tag{18}$$

and

$$T_l = T_r = e^{-ik\xi(r)L}\frac{(1-\mathrm{R})e^{ik\xi(r)nL}}{1 - \mathrm{R}e^{2ik\xi(r)nL}}, \tag{19}$$

where $\mathrm{R} := (\frac{n-1}{n+1})^2$. Expressing the refractive index of the slab as $n = \eta + i\kappa$, substituting it into Eqs. (18) and (19), and multiplying them by their complex conjugates give the reflection and transmission coefficients as follows

$$\mathcal{R} = \frac{|\mathrm{R}|[(1-G)^2 + 4G\sin^2\delta/2]}{(1-G|\mathrm{R}|)^2 + 4G|\mathrm{R}|\sin^2(\delta+\phi)/2} \tag{20}$$

$$\mathcal{T} = \frac{|1-\mathrm{R}|^2 G}{(1-G|\mathrm{R}|)^2 + 4G|\mathrm{R}|\sin^2(\delta+\phi)/2}, \tag{21}$$

where

$$|\mathrm{R}| := \frac{(\eta-1)^2+\kappa^2}{(\eta+1)^2+\kappa^2}, \quad G := e^{-2k\xi(r)\kappa L}, \quad \delta := 2k\xi(r)\eta L, \quad \phi := 2\tan^{-1}(\frac{2\kappa}{\eta^2+\kappa^2-1}). \tag{22}$$

The spectral singularities correspond to the real and positive values of the wavenumber $k$ for which $M_{22} = 0$ [1]. This implies that the reflection and transmission amplitudes diverge. So it occurs when

$$e^{-2ik\xi(r)nL} = \mathrm{R}. \tag{23}$$

Taking the natural logarithm of both sides of this equation yields

$$k\xi(r)L = -\frac{1}{2in}\ln\mathrm{R}, \tag{24}$$

and in follow by inserting $n = \eta + i\kappa$ in Eq. (24), we get

$$k\xi(r)L = -\frac{1}{2i(\eta + i\kappa)} \ln\left(\frac{\eta - 1 + i\kappa}{\eta + 1 + i\kappa}\right)^2 \tag{25}$$

Because $k\xi(r)L$ is real, therefore it equates with the real part of the right-hand side of this equation and so the imaginary part of the right-hand side is set to zero. This gives

$$k\xi(r)L = \frac{1}{2\kappa} \ln|\mathrm{R}| \tag{26}$$

and

$$\eta \ln|\mathrm{R}| + \kappa(\phi - 2m\pi) = 0, \tag{27}$$

where $m$ is an arbitrary integer.

## 4. Threshold gain

The gain coefficient of an active medium is defined by [32]

$$g := -\frac{4\pi\kappa}{\lambda}, \tag{28}$$

where $\kappa$ is imaginary part of the refractive index $n$ and $\lambda := 2\pi/k$ is the wavelength. The value of gain coefficient for which the spectral singularity takes place is called threshold gain, which is the lasing threshold condition. For getting the threshold gain of the active medium, we derive $\kappa$ from Eq. (26)and then substitute it in Eq. (28) which yields

$$g = g_{th} := \frac{1}{2[1 + \frac{r^2}{2z_0^2}]L} \ln\left[\frac{(\eta+1)^2 + \kappa^2}{(\eta-1)^2 + \kappa^2}\right], \tag{29}$$

where we have used Eq. (12) for $\xi(r)$.

## 5. Accounting for dispersion

For a more realistic situation we take into account the effect of dispersion, i.e., the refractive index $n$ and so $\eta$ and $\kappa$ are not independent from the frequency. In this regard, suppose that the slab contains a doped host medium of refraction index $n_0$ that we can model by a two-level atomic system with lower and upper level population densities $N_l$ and $N_u$, resonance frequency $\omega_0$, damping coefficient $\gamma$, and the dispersion relation [33]

$$n^2 = n_0^2 + \frac{\omega_p^2}{\omega_0^2 - \omega^2 - i\gamma\omega}, \tag{30}$$

where $\varepsilon_0$ is the permeability of the vacuum, $\omega_p^2 := \frac{(N_l - N_u)e^2}{m_e \varepsilon_0}$, and $e$ and $m_e$ are electron's charge and mass, respectively. Substituting the refractive index of the slab $n = \eta + i\kappa$ into Eq. (30) and taking $\kappa = \kappa_0$ at $\omega = \omega_0$, we get the following approximate equations for the real and imaginary parts of the refractive index

$$\eta \simeq n_0 - \kappa_0 F_1 \tag{31}$$

and

$$\kappa \simeq \kappa_0 F_2, \tag{32}$$

where

$$F_1 := \frac{(1 - \hat{\omega}^2)\hat{\gamma}}{(1 - \hat{\omega}^2)^2 + \hat{\gamma}^2\hat{\omega}^2}, \quad F_2 := \frac{\hat{\omega}\hat{\gamma}^2}{(1 - \hat{\omega}^2)^2 + \hat{\gamma}^2\hat{\omega}^2}, \tag{33}$$

$\hat{\omega} = \frac{\omega}{\omega_0}$, and $\hat{\gamma} = \frac{\gamma}{\omega_0}$.
Next, by inserting Eqs. (31) and (32) in Eq. (26) and using Eqs. (12) and (8), we find

$$\kappa_0 \simeq \frac{\frac{1}{2}\ln \mathrm{R}_0}{\frac{2F_1}{n_0^2-1} + [2\pi(\frac{\hat{\omega}}{\lambda_0}) + \frac{r^2}{16\pi(f/D)^4}(\frac{\hat{\omega}}{\lambda_0})^3]LF_2}. \tag{34}$$

where $\mathrm{R}_0 = (\frac{n_0-1}{n_0+1})^2$. By this way, threshold gain is obtained from Eq. (28) as follows

$$g_{th} := -4\pi\kappa_0/\lambda_0, \tag{35}$$

where $\lambda_0 := 2\pi c/\omega_0$, and $\kappa_0$ is evaluated from Eq. (34). In deriving Eqs. (31)-(35) the quadratic- and higher-order terms of $\kappa_0$ are neglected. Using the same approximation to calculate the reflection and transmission coefficients by Eqs. (20) and (21), we can use the following relations instead of Eq. (22)

$$|\mathrm{R}| \simeq \mathrm{R}_0(1 + \frac{4\kappa_0 F_1}{n_0^2-1}), \quad G \simeq e^{-2k\xi(r)\kappa_0 LF_2}, \quad \delta \simeq 2k\xi(r)L(n_0 - \kappa_0 F_1), \quad \phi \simeq \frac{4\kappa_0 F_2}{n_0^2-1}. \tag{36}$$

Then, extracting $|\mathrm{R}|$ from Eq. (27), substituting in Eq. (26), and using Eqs. (8), (12) and (31)-(33), we find the following mode equation

$$\frac{r^2 n_0 L}{16\pi(f/D)^4\lambda_0^3}\hat{\omega}^4 + [\frac{2\pi n_0 L}{\lambda_0} + \frac{\ln \mathrm{R}_0}{2\hat{\gamma}^2}]\hat{\omega}^2 - \pi m\hat{\omega} - \frac{\ln \mathrm{R}_0}{2\hat{\gamma}^2} = 0. \tag{37}$$

The solutions of this equation are the resonance frequencies related to the spectral singularities which, in addition to depending on the physical characteristics of the active medium, they are $r$-dependent and change with the relative aperture of lens

## 6. Numerical results

In order check the numerical results of the theoretical findings in the previous sections, we consider a slab composed of a semiconductor gain medium with the specifications: $n_0 = 3.4$, $L = 300\mu m$, $\lambda_0 = 1500nm$, and $\hat{\gamma} = 0.02$, in the back focal plane of a thin convergent lens with variable relative aperture under illumination of a Gaussian beam. The reflection and transmission coefficients from the slab are calculated from Eqs. (20) and (21), using Eq. (36). Fig. 2 illustrates the logarithmic plots of the reflection (dashed curve) and transmission (full curve) coefficients of a light at center of the focused Gaussian ($r = 0$) from the specified slab versus the wavelength. The plots show the zero-width resonance frequencies with the free spectral range is of the order of $\delta\lambda \approx 1nm$.

To find out how the spectral singularities and the corresponding resonance frequencies depend to the transverse distance $r$, we calculate the reflection and transmission coefficients at various transverse distances on the beam spot. Fig. 3 shows the logarithmic curves of the calculated transmission coefficient at various transverse distances. Based on these plots, the spectral singularities corresponding resonances are $r$-dependent, and as $r$ increases, shift to the higher wavelengths. In fact, in a Gaussian beam the outer rays, in comparison to the central ray, are obliquely propagating, that is, the incident angle of the rays striking to the front face of slab increases, radially and so the optical path lengths experienced by the rays in passing through the slab increase, radially. Recently, it has been illustrated that the resonance frequencies related to spectral singularities depend strongly on the incident angle [17, 20].

Now, we study the dependence of the threshold gain on the wavelength at different transverse distances from the center of the beam. For this purpose, we calculate the threshold gain from Eq. (34) using Eq. (35). Fig. 4. shows the plots of the calculated threshold gain versus the wavelength at various transverse distances from the center of the beam. Upon these plots the threshold gain of the slab becomes maximum at the center of the beam and decreases as $r$ increases.. This is due to that the amplitude of the field distribution across the Gaussian beam decreases radially according to Eq. (6).

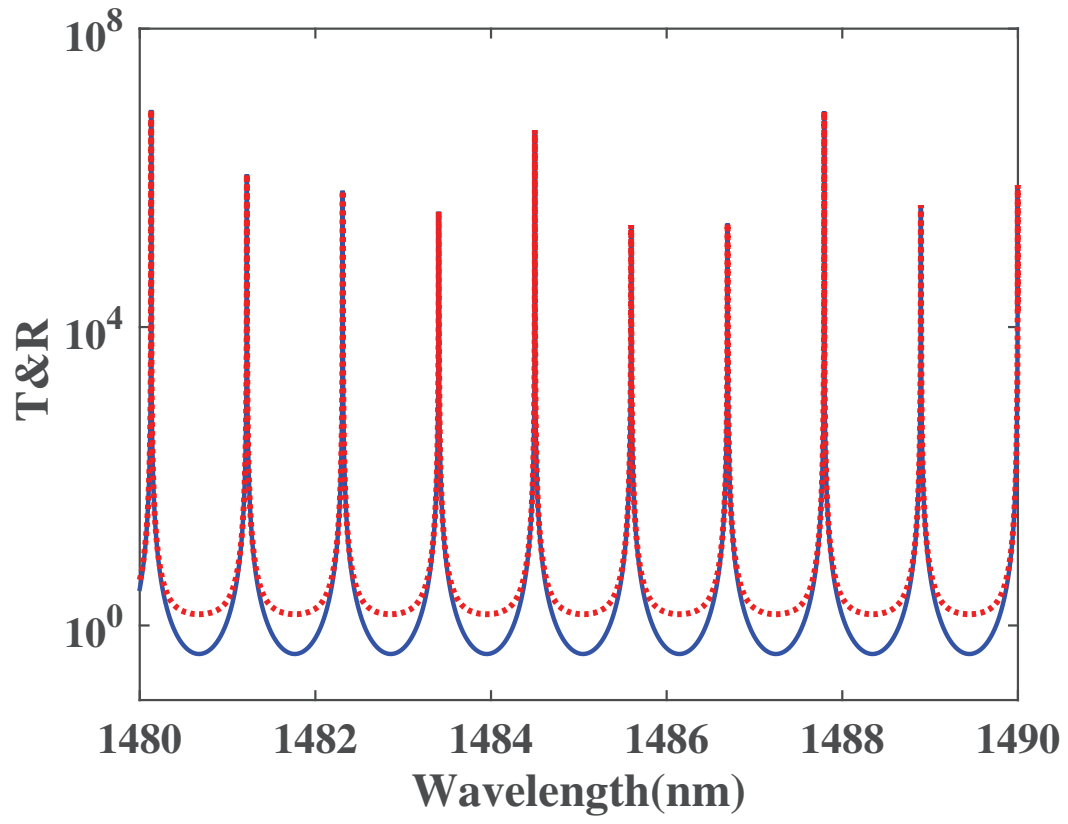

Fig. 2. Logarithmic plots of the reflection (dashed curve) and transmission (full curve) coefficients versus the wavelength of a Gaussian beam from a slab composed of a semiconductor gain medium with the specifications $n_0 = 3.4$, $L = 300\mu m$, $\lambda_0 = 1500nm$, and $\hat{\gamma} = 0.02$ in focus of a thin lens with the relative aperture of $f/D = 15$ that have been calculated using Eqs. (20), (21) and (36) at center of the beam spot.

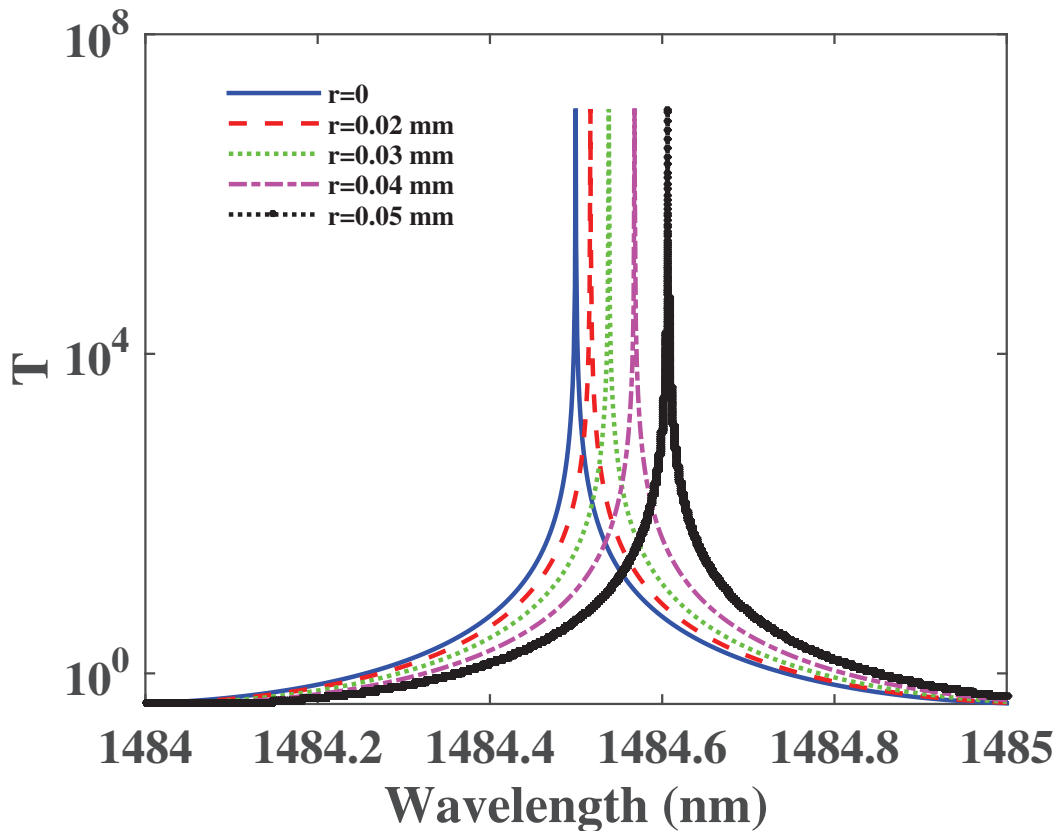

Fig. 3. Logarithmic plots of the transmission coefficient versus the wavelength of a Gaussian beam from a slab composed of a semiconductor gain medium with the specifications $n_0 = 3.4$, $L = 300\mu m$, $\lambda_0 = 1500nm$, and $\hat{\gamma} = 0.02$ in focus of a thin lens with the relative aperture of $f/D = 15$ that have been calculated using Eqs. (21) and (36) at various transverse distances from the center of the beam spot.

In the proposed configuration, the relative aperture of the thin lens $f/D$ is a factor that influences the threshold gain of the active medium and the corresponding spectral singularities, according to Eqs. (34), (35), and (37). Therefore, to consider this effect we can use various lenses with different relative apertures. We calculated the transmission coefficient of a Gaussian beam from the specified slab in focus of the lenses with different relative apertures have been calculated using Eqs. (21) and (36) and plotted them, as shown in Fig. 5. According to these plots, the use of the lenses with higher relative apertures leads shifting the spectral singularities to the lower wavelengths (higher frequencies). One can use this ability to tune the spectral singularities. Also, Fig. 6 shows the plots of the threshold gain calculated from Eqs. (34) and (35) for the specified slab in focus of the lenses with different relative apertures. These plots illustrate that using the thin lenses with higher relative apertures, the threshold gain of the slab increases.

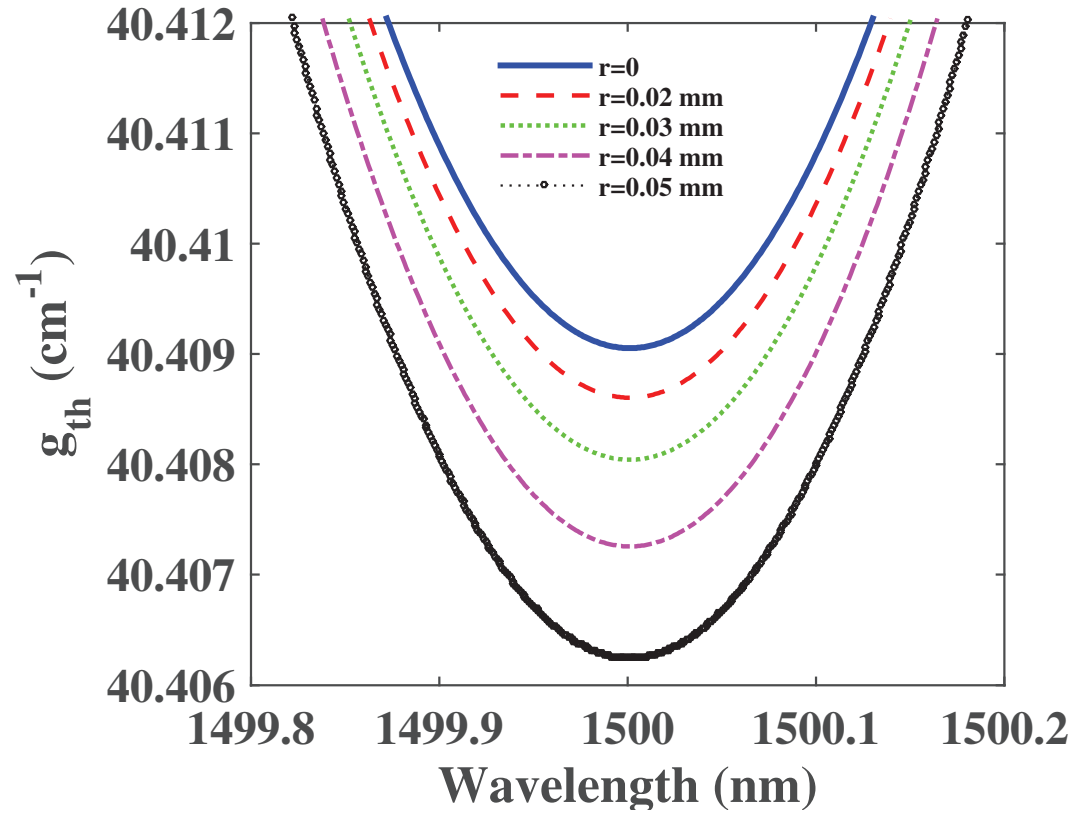

Fig. 4. The plots of the threshold gain versus the wavelength of a Gaussian beam for a slab composed of a semiconductor gain medium with the specifications $n_0 = 3.4$, $L = 300\mu m$, $\lambda_0 = 1500nm$, and $\hat{\gamma} = 0.02$ in focus of a thin lens with the relative aperture of $f/D = 15$ that have been calculated using Eqs. (34) and (35) at various transverse distances from the center of the beam spot.

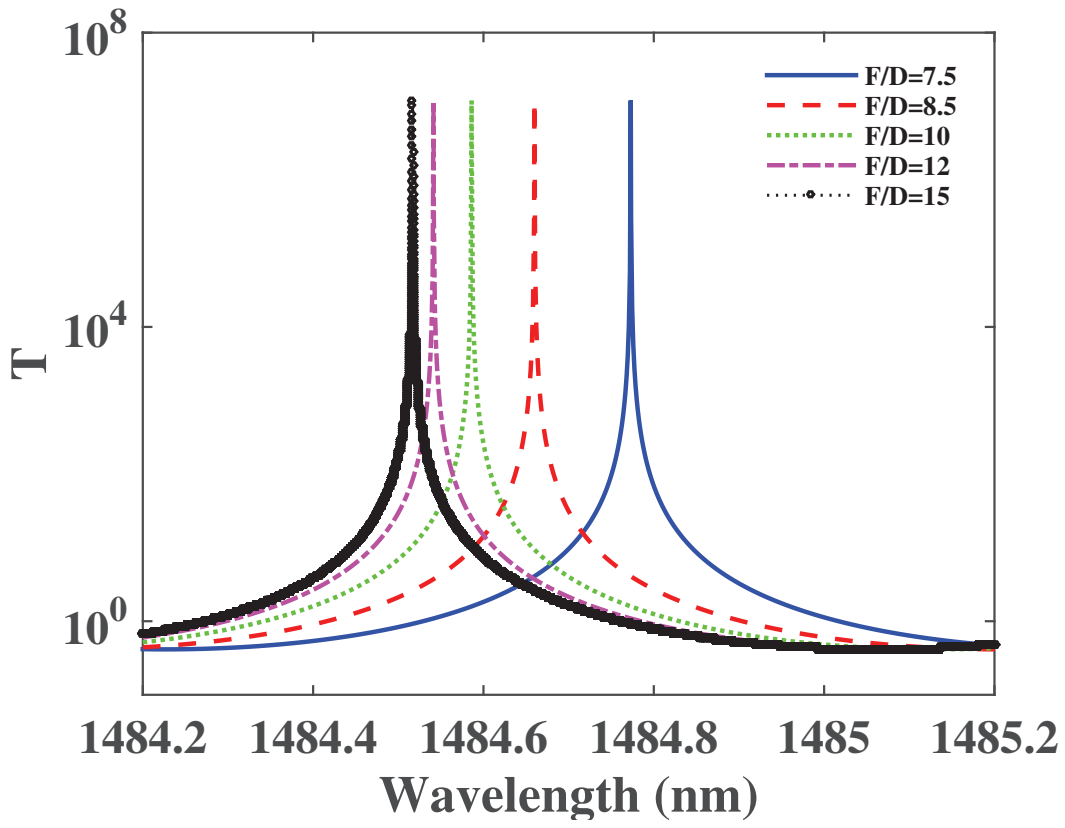

Fig. 5. Logarithmic plots of the transmission coefficient versus the wavelength of a Gaussian beam from a slab composed of a semiconductor gain medium with the specifications $n_0 = 3.4$, $L = 300\mu m$, $\lambda_0 = 1500nm$, and $\hat{\gamma} = 0.02$ in focus of the thin lenses with different relative apertures that have been calculated using Eqs. (21) and (36) at $r = 0.02mm$ from the center of the beam spot.

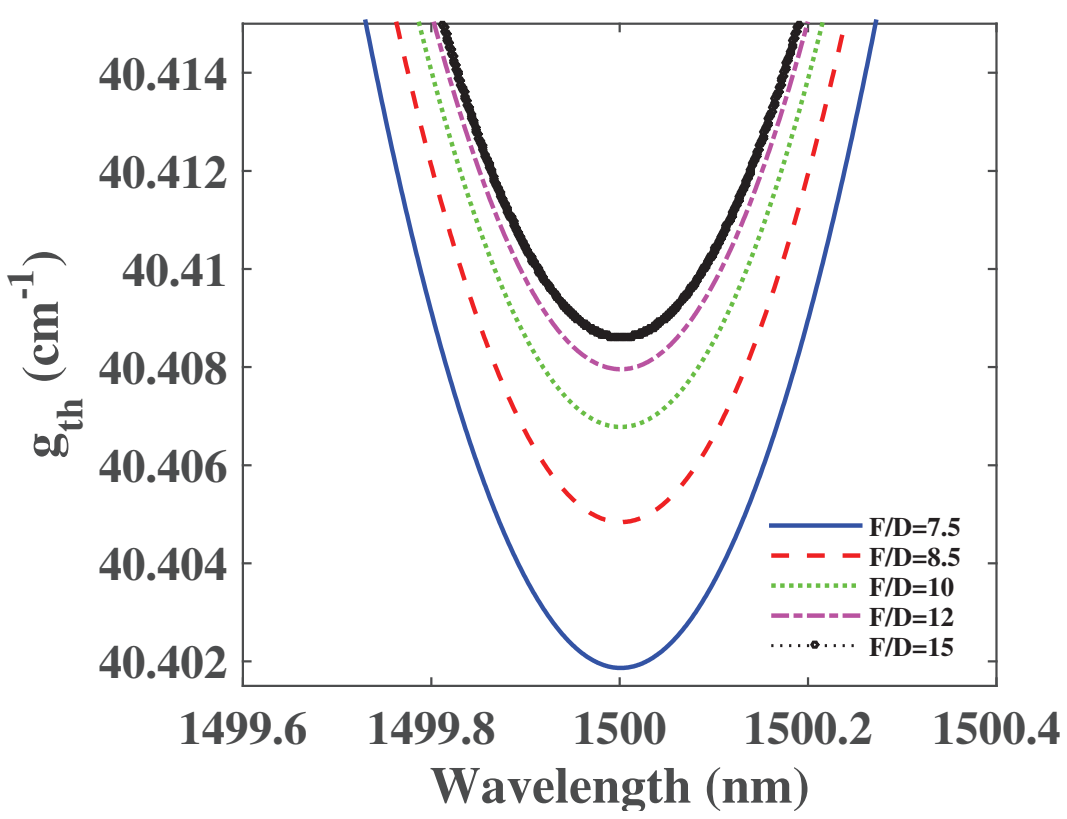

Fig. 6. The plots of the threshold gain versus the wavelength of a Gaussian beam for a slab composed of a semiconductor gain medium with the specifications $n_0 = 3.4$, $L = 300\mu m$, $\lambda_0 = 1500nm$, and $\hat{\gamma} = 0.02$ in focus of the lenses with different relative apertures that have been calculated using Eqs. (34) and (35) at $r = 0.02mm$ from the center of the beam spot.

## 7. Conclusion

We studied the spectral singularities and the threshold gain of a laser slab in focus of a thin lens under illumination of a Gaussian beam. Our purposes of this study are that, first, to extend the studies done on the spectral singularities of the gain medium for the Gaussian beams, and second, tuning the spectral singularities corresponding resonances by use of the focusing elements of the configuration. The novel and worth-mentioning features of this study are expressed as follows:
1-The field distribution of a Gaussian beam in focus of a thin lens is approximated by a plane wave which it's phase and amplitude are $r$-dependent, according to Eq. (10). Physically, in a Gaussian beam the outer rays, in comparison to the central ray, are obliquely propagating, therefore the Gaussian beam in focus can be described as a set of plane waves which their's wave vectors make larger angle to the optical axis, as $r$ increases.
2- The transfer matrix for a slab composed of optically active material under illumination of a focused Gaussian beam is obtained in Eq. (15) which is $r$-dependent.
3- Because the amplitude of the field distribution across the Gaussian beam decreases radially, according to Eq. (6), therefore the corresponding threshold gain of the medium also decreases radially, as illustrated in Fig. 4.
4- In a Gaussian beam the rays that propagate through the slab experience larger optical path length by moving away from the center of the beam, therefore the corresponding spectral singularities take place at the higher wavelengthes, as shown in Fig. 3.
5- Threshold gain and spectral singularity vary with the relative aperture, $f/D$, of the focusing lens, according to Eqs. (35) and (37). Increasing the relative aperture leads to shift the spectral singularities corresponding resonances to the lower wavelengths and to increase of the corresponding threshold gain of the active medium, as illustrated in Figs. 5 and 6.
This study can be extended to the other features of the electromagnetic waves in the gain and loss mediums such as coherent perfect absorption and invisibility. Also, this study can be done experimentally by providing highly sensitive intensity detectors.

## 8. Disclosures

The authors declare no conflicts of interest.